\documentclass[11pt]{article}

\usepackage[final]{acl}

\usepackage{times}
\usepackage{latexsym}

\usepackage[T1]{fontenc}

\usepackage[utf8]{inputenc}

\usepackage{microtype}

\usepackage{inconsolata}

\usepackage{graphicx}

\usepackage{booktabs}
\usepackage{multirow}
\usepackage{array}
\usepackage{graphicx}
\usepackage{geometry}
\usepackage{colortbl}
\usepackage{xcolor}
\usepackage{makecell}
\usepackage{amsmath}
\usepackage{amsfonts}
\usepackage{amssymb}
\usepackage{svg}
\title{Yet Even Less Is Even Better For Agentic, Reasoning, and Coding LLMs}

\author{
\hyperref[sec:contributors]{CodeArts Model Team}
}

\begin{document}
\maketitle
\begin{abstract}
Training effective software engineering agents requires large volumes of task-specific trajectories, incurring substantial data construction costs.
Inspired by the "Less-Is-More" hypothesis in mathematical reasoning, we investigate its extension to agentic scenarios and propose an end-to-end training framework that achieves superior agentic capabilities with fewer but higher-quality training trajectories. 
This is achieved via STITCH (Sliding-memory Trajectory Inference and Task Chunking Heuristic), a coarse-to-fine mechanism that filters low-value noise and retains decision-critical tokens to maximize training signal quality.
We conduct experiments across multiple agent frameworks(e.g., mini-SWE-agent, MSWE-agent), model scales (30B to 355B), and multilingual settings (Python, Java, and ArkTS). On SWE-bench Verified, models trained with STITCH achieve up to 63.16\% relative improvement over base models.
On Multi-SWE-bench (Java), MiniMax-M2.5-STITCH achieves 43.75\% with our CodeArts Agent scaffold (+16.67\%). On HarmonyOS (ArkTS), GLM-4.7-STITCH improves the compilation pass rate to 61.31\% (+43.34\%) with less than 1K training trajectories. Our results confirm that the "Less-Is-More" paradigm generalizes effectively to complex agentic tasks across diverse languages and model scales.
\end{abstract}

\section{Introduction}
Recently, using large language models (LLMs) as the core to build agents for solving complex tasks has emerged as a prominent trend in artificial intelligence~\cite{wang2024survey, xi2025rise, yao2023react}. Such tasks demand a diverse set of capabilities --- encompassing \textit{long-horizon reasoning}, \textit{code generation}, \textit{tool use}, and \textit{multi-turn interaction} --- collectively referred to as \textbf{Agentic abilities}~\cite{wu2025masksearch, fang2025towards}. In the software engineering domain, SWE-bench has become a canonical benchmark for evaluating these abilities~\cite{SWE-bench, liu2024deepseek, comanici2025gemini, anthropic2024swebench}, where agents are required to autonomously resolve real-world GitHub issues by understanding natural language requirements, navigating large codebases, modifying multiple files, executing tests, and iteratively refining solutions until success.


A substantial body of research has focused on enhancing LLM' Agentic abilities on SWE-bench~\cite{swe-factory, lin2025se, swe-gym, swe-lego}. Among these, training-based methods~\cite{swe-gym, r2e-gym, swe-lego, swe-master} have proposed various methodologies including supervised fine-tuning (SFT), reinforcement learning (RL), and their combinations, achieving notable improvements. However, they share several limitations: (1) \textbf{High data construction costs}. These methods typically require large-scale task construction through human-written issues and test cases~\cite{swe-gym}, or complex synthetic data generation pipelines, incurring substantial annotation overhead. (2) \textbf{Lack of investigation on data efficiency}. Prior works often focus on scaling data quantity without thoroughly exploring the relationship between data quality and training effectiveness. (3) \textbf{Limited exploration of large-scale models}. Most experiments are conducted on small to medium-sized models (e.g., 7B-72B parameters), leaving the training dynamics of larger models (e.g., 100B+) underexplored.


Inspired by LIMO \cite{limo}, which proposes the "Less-Is-More Reasoning" hypothesis in the context of pure mathematical reasoning and demonstrates that sophisticated reasoning can emerge with only a few hundred carefully curated examples, we ask a compelling question: \textit{Can the "Less-is-More" paradigm be extended from pure reasoning to Coding and Agentic scenarios, particularly across models of varying scales?}


In this paper, we propose a Less-is-more-style training framework for SWE agents, aiming to achieve stronger Agentic capabilities with less but higher-quality training data. Our core contributions are outlined as follows:

\begin{enumerate}
    \item \textbf{Theoretical Extension.} We extend "Less-Is-More" hypothesis from pure reasoning tasks to Agentic scenarios. We demonstrate that when code understanding and tool use capabilities have been adequately encoded in foundation models, Agentic abilities can also be efficiently elicited through carefully curated high-quality trajectories.

    \item \textbf{End-to-End Pipeline.} We construct a complete end-to-end training pipeline, including (1) task construction from real-world data; (2) trajectory-based step-level training data curation; (3) model fine-tuning; and (4) automated evaluation. This pipeline scales to multi-language scenarios and reduces dependence on human annotations.

    \item \textbf{Step-level Trajectory Analysis Tool.} We design a step-level trajectory analysis tool that combines heuristic rules with LLM agents to identify high-value tokens, i.e., tokens containing critical reasoning, decision-making, or code fix operations. By filtering low-value noise and enhancing high-value signals, the training effect is significantly improved.

    \item \textbf{Systematic Experiments.} We conduct comprehensive experiments across multiple SWE agent frameworks (e.g., MSWE-agent, OpenCode), models of various scales \cite{qwen3, glm45, minimax2026m25} (including large models with 100B+ parameters), and multi-lingual (Python/Java/Arkts) scenarios, demonstrating the effectiveness of the proposed framework and training methodology.
\end{enumerate}


Experimental results show that our method achieves significant improvements over baseline models across multiple benchmarks and languages. On SWE-bench Verified, models trained with STITCH achieve up to 63.16\% relative improvement over base models. On Multi-SWE-bench (Java), MiniMax-M2.5-STITCH achieves 43.75\% with our CodeArts Agent scaffold (+16.67\%), reaching state-of-the-art performance among open-source models. On HarmonyOS (ArkTS), GLM-4.7-STITCH improves the compilation pass rate to 61.31\% (+43.34\%). These results consistently demonstrate the scalability and practicality of the ``Less-is-More'' paradigm across diverse Agentic scenarios.

\section{Related Work}
\subsection{Data Construction for Code Agent}

High-quality training data is fundamental to code agent development. Recent efforts shift from single-turn code generation~\cite{OpenCodeInstruct} to multi-turn, repository-level trajectories with executable feedback~\cite{swe-gym}. SWE-Gym ~\cite{swe-gym} pioneers this direction with real-world task instances, demonstrating that supervised fine-tuning on collected trajectories yields substantial improvements. Subsequent work explores scalable data synthesis: R2E-Gym~\cite{r2e-gym} introduces backtranslation-based generation achieving strong performance on SWE-bench Verified; SWE-smith~\cite{swe-smith} automates bug synthesis to create large-scale instances from diverse repositories; SWE-Factory~\cite{swe-factory} and SWE-rebench V2~\cite{swe-rebench-v2} build language-agnostic pipelines supporting tasks across multiple programming languages. Beyond supervised paradigms, Self-Play SWE-RL \cite{self-play-swe-rl} achieves notable improvement through autonomous bug injection and repair without human annotation, while SWE-World \cite{swe-world} proposes replacing physical execution with learned surrogate models for scalable training.

\subsection{Training Paradigm for Code Agent}

Training code agents primarily relies on two complementary approaches: supervised fine-tuning (SFT) and reinforcement learning (RL). SFT leverages expert demonstrations to distill capabilities from stronger models. SWE-Gym \cite{swe-gym} establishes the viability of trajectory-based SFT, while SWE-Lego \cite{swe-lego} introduces refinements including step-level error masking, which excludes erroneous tokens from loss computation, and curriculum learning that progressively increases task difficulty, achieving strong results on SWE-bench-Verified using SFT alone. RL enables learning from environmental feedback without requiring expert trajectories. SWE-Master \cite{swe-master} demonstrates that combining SFT initialization with RL fine-tuning yields superior performance compared to either method in isolation, highlighting the complementary nature of these paradigms.

\section{Approach}
\subsection{Overview}

\begin{figure*}[t]
    \centering
    \includegraphics[width=\textwidth]{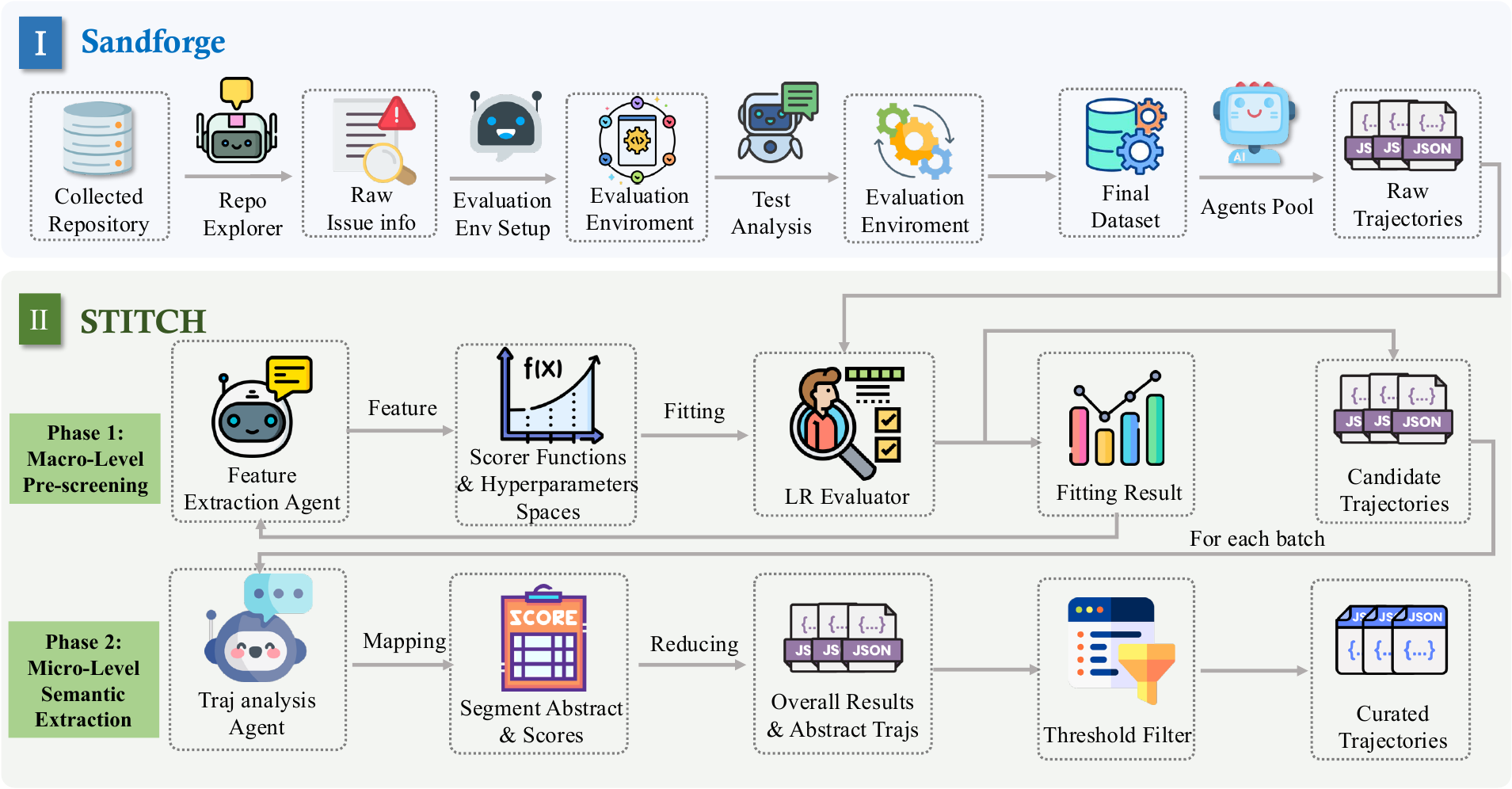}
    \caption{Overview of our proposed framework. SandForge converts GitHub issues into executable tasks and collects raw trajectories. STITCH filters trajectories through macro-level pre-screening and micro-level semantic analysis using a Map-Reduce paradigm with sliding memory mechanism.}
    \label{fig:overview}
\end{figure*}

We propose a ``Less-Is-More'' training framework that prioritizes training signal quality over data volume. As shown in Figure~\ref{fig:overview}, the framework consists of two main components: SandForge and STITCH. SandForge first converts real-world GitHub issues into standardized executable tasks, retaining trajectories, reward signals, and verifier outputs as structured training artifacts. Multiple agent frameworks are then deployed to collect raw trajectories. STITCH subsequently filters these trajectories in two stages: a Logistic Regression model performs macro-level pre-screening based on automatically discovered agentic features, and the remaining trajectories undergo micro-level semantic analysis. Since agent trajectories are often too long to fit within the context window of an LLM evaluator, chunking is necessary. However, naive fixed-length chunking severs action-observation dependencies at segment boundaries, leading to distorted quality assessments. To address this, STITCH employs a Map-Reduce paradigm with a sliding memory mechanism: semantically intact chunk boundaries are selected via heuristic safe-split points, and in the Map phase, a compressed memory state from the preceding segment is carried forward to maintain cross-segment semantic coherence without exceeding the context window. In the Reduce phase, local segment scores are aggregated into a global trajectory quality assessment, enabling the extraction of high-value sub-task segments even from globally suboptimal trajectories and maximizing training data utilization.

\subsection{Unified Data Construction and Evaluation Framework}

The SWE tasks require agents to manipulate real software artifacts under constrained execution environments, where a task includes not only instructions but also repository states, environment dependencies, test scripts, permissions, and verification semantics. As a result, the central challenge is how to execute a complex software process under a unified abstraction. 

We further argue that coding agent infrastructure should treat every execution as a potential data-construction event. A full task execution includes instructions, environment context, agent behavior sequences, tool calls, observations, rewards, exception information, and generated patches. These signals can be preserved as structured trajectories, rewards, patches, and artifacts rather than remaining as scattered logs. The same execution can support evaluation, diagnosis, and training. This perspective has direct implications for both Fine-tuning and RL. Structured dialogues and tool-use traces can become supervised data; retained rewards and rollouts can support offline RL or preference modeling; verifier outputs and patch differences can support failure analysis and dataset filtering. Importantly, this data-construction capability does not presuppose that tasks come only from public benchmarks. The same execution-and-retention mechanism can operate over both benchmark suites and self-constructed datasets.

To support this setting, we implement \textbf{SandForge}, an internal end-to-end framework for sandbox-based coding-agent data construction and evaluation. In practice, the framework is designed to support heterogeneous sandbox-based datasets, CLI agents, and LLM backends under a common execution interface. Its methodology combines two tightly coupled stages: upstream task construction from real software repair records, and downstream unified execution and evaluation with structured artifact retention.

\subsubsection{Task Construction from Real Software Repair Records}

High-quality coding-agent evaluation begins with realistic and executable tasks. In repository-centric settings, task instances are engineered representations of real repair events rather than naturally given inputs. Our pipeline therefore starts from GitHub repositories and collects repair-related metadata from issues, pull requests, commits, and tests.

Candidate events are filtered to retain bug-fix-oriented cases with traceable issue-PR links and informative patch and test changes. The retained events are then converted into standardized task instances with explicit repository states, environment specifications, test assets, verification entry points, and executable environment or image build logic.

Each candidate instance is further validated through staged executability checks: the clean baseline must run, the test patch must expose the defect, and the original fix must make the relevant tests pass. Only instances satisfying these constraints are retained. This process yields tasks that are both realistic in provenance and directly usable for downstream agent execution and trajectory collection.

\subsubsection{Unified Runtime and Multi-Agent Adaptation}

The \textbf{SandForge} framework is organized around five core abstractions: task specification, execution environment, agent adaptation, single execution instance, and experiment-batch orchestration. Together they define what should be solved, where execution occurs, how agents interact with the environment, what constitutes one run, and how runs are organized into reproducible experiments.

The runtime is built on the decoupling of task semantics, execution environments, and agent behavior. Each execution instance follows an explicit lifecycle: environment setup, agent setup, agent execution, patch export, verifier execution, artifact collection, and result finalization. At the batch level, the framework preserves configurations, timeout policies, artifact paths, and partial results for resumption and analysis.

Heterogeneous agents are integrated through a unified adaptation layer. Instead of relying on benchmark-specific wrappers, the framework reduces each agent to the same contract: setup, execution, patch production, and result retention. Runtime details such as model names, provider mappings, sampling parameters, environment variables, proxies, certificates, and optional extensions are normalized through a common path, which is essential for reproducibility and portability in mixed-provider and restricted-network environments. To remain faithful to benchmark semantics, the runtime also preserves benchmark-facing artifacts through explicit verifier integration, patch export, and replay-aware result retention, including baseline-sensitive patch handling and replay-consumable artifact management, rather than relying on local execution success alone, since the final working tree may differ from the artifact consumed by official replay.

In internal deployments, the framework has already been instantiated across multiple coding agents, including \texttt{mini-swe-agent}~\cite{minisweagent}, \texttt{Claude Code}~\cite{anthropic2024claudecode}, \texttt{gemini-cli}~\cite{google2025geminicli}, \texttt{OpenCode}~\cite{opencode}, \texttt{MSWE-agent}~\cite{zan2024swe}, and our internal~\texttt{CodeArts-Agent}. The experiments in this paper focus on a narrower subset of agents and benchmarks, as described in the experimental setup.

\subsubsection{Execution Outputs as Reusable Data}

The output of \textbf{SandForge} is not limited to benchmark scores. For each execution instance, the framework retains trajectories, tool traces, verifier outputs, rewards, patches, artifacts, and execution metadata. These outputs support several downstream uses: trajectories for fine-tuning, rewards and rollouts for offline reinforcement learning or preference modeling, and patches plus verifier logs for failure analysis and dataset filtering.

For each trajectory record, downstream consumers can inspect associated metadata to determine whether the run passed the verifier, whether the target tests succeeded, and whether execution-level anomalies such as tool failures, runtime exceptions, or environment errors occurred. This makes it possible to perform quality filtering, error-aware stratification, and selective data retention. 

For HarmonyOS (ArkTS) scenarios, we employ a special two-stage strategy for trajectory labeling, since many tasks involve UI-related behaviors that cannot be reliably evaluated through code-level test cases alone. First, we filter out the trajectories that failed at compilation. Second, candidate trajectories that passed compilation are further validated through an automated visual preview pipeline. Each compiled project is deployed to a device via the HarmonyOS HDC toolchain, and runtime screenshots are captured across all identifiable UI pages. A multimodal language model (Qwen/Qwen3.5-122B-A10B) then evaluates whether the rendered interface aligns with the original requirement description, filtering out trajectories that compile successfully but fail to implement the specified functionality.

As a result, the same framework supports both evaluation and training-data accumulation. This dual role is the main reason we treat \textbf{SandForge} as an end-to-end data construction and evaluation framework rather than merely a benchmark runner. 

\subsection{Coarse-to-Fine Trajectory Data Curation}
To address the prohibitive costs and hallucination risks in curating massive agent trajectories, we propose \textbf{STITCH} (\textbf{S}liding-memory \textbf{T}rajectory \textbf{I}nference and \textbf{T}ask \textbf{C}hunking \textbf{H}euristic), a novel coarse-to-fine trajectory data curation framework.
STITCH operates in two sequential stages:

\begin{itemize}
    \item \textbf{Macro-Level Pre-screening via Statistical Features} Employs a lightweight Logistic Regression model to rapidly filter out statistically suboptimal trajectories based on empirically derived features and automatically discovered features by a feature extraction and fitting agent.
    \item \textbf{Micro-Level Semantic Extraction and Verification} For the promising candidates, deploys a micro-level trajectory semantic segmentation and evaluation agent, to verify the logical coherence within a segment and extract local segments with high quality for training.
\end{itemize}

By cascading statistical efficiency with neural semantic rigor, STITCH successfully synthesizes high-quality datasets optimal for SFT or any other downstream alignments


\begin{figure}[h!]
    \centering 
\includegraphics[width=0.5\textwidth]{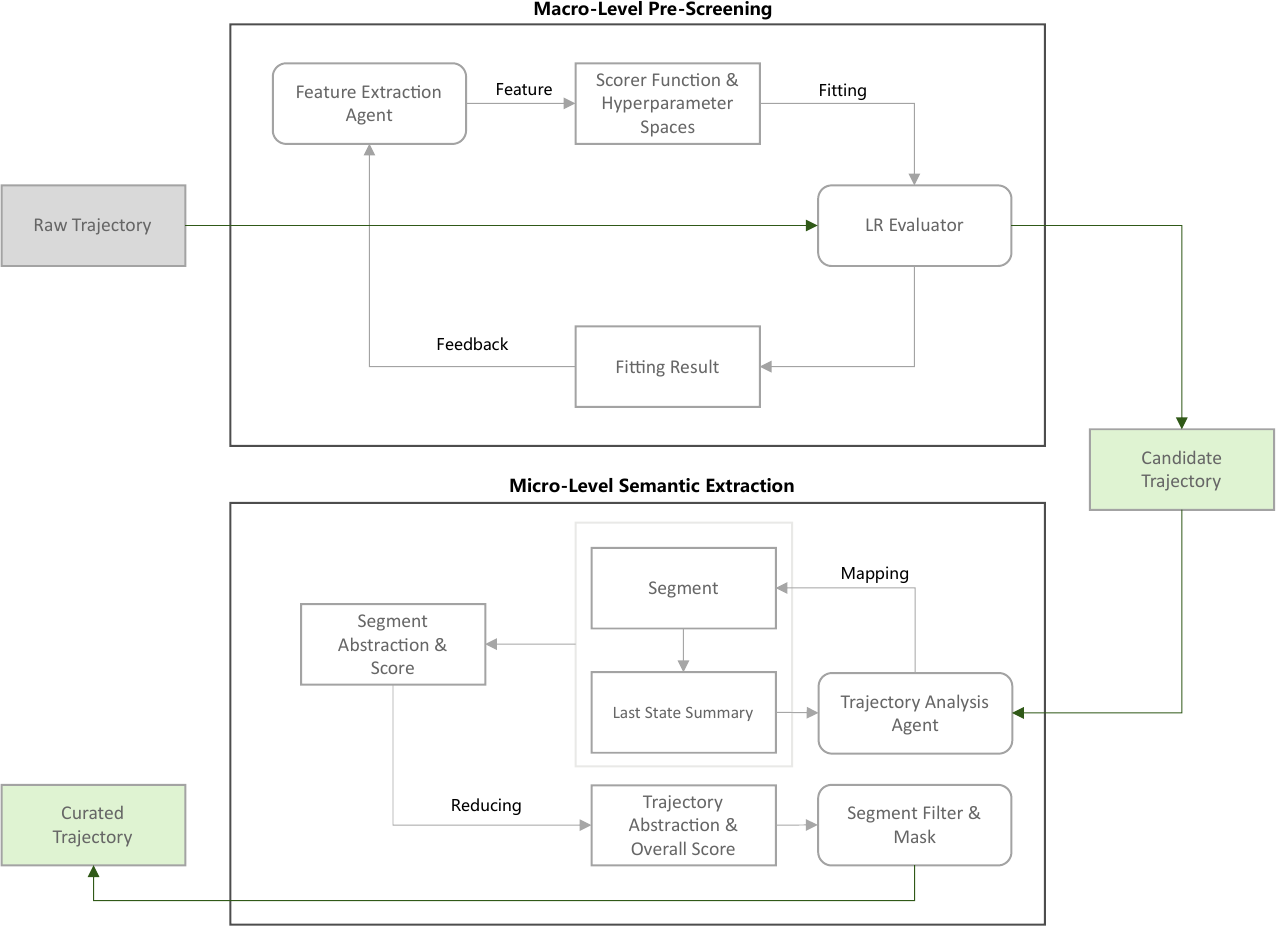} 
    \caption{Two-Stage Trajectory Curation Pipeline of STITCH}
    \label{fig:STITCH} 
\end{figure}

\subsection{Automated Feature Discovery and Weight Optimization} 

Compared to ordinary multi-turn conversations, trajectory data contain more information about the actions an agent takes in the context of a specific task. To systematically quantify the quality of an agent's trajectory $\mathcal{T}$ in a specific scenario $\mathcal{S}$, we define the total trajectory score $S_{total}$ as a linear combination of $k$ active scoring functions $f_i \in \mathcal{F}^{(\mathcal{S})}$. Each function is parameterized by a scenario-specific hyperparameter set $\Theta^{(\mathcal{S})} = \{\theta_1, \dots, \theta_k\}$:

\begin{equation}S_{total}(\mathcal{T}, \mathcal{S}) = \sum_{i=1}^{k} f_i(\mathcal{T}; \theta_i)\end{equation}

To decouple mathematical logic from domain-specific heuristics, we abstract the initial scoring function $f_i$ into three general function families for the fitting target of the Feature Extraction Agent in the cold start phase:

1. \textbf{Bounded Linear Reward}
Applied to cumulative metrics such as code production volume or tool diversity. To prevent score exploitation (e.g., verbose but redundant generation), the reward grows linearly but is strictly capped. Given an extracted feature $x(\mathcal{T})$, weight $w$, and maximum score $M$:
\begin{equation}
f_{cap}(x; w, M) = \min \big( w \cdot x(\mathcal{T}), \; M \big)
\end{equation}

2. \textbf{Proportional Reward}
Utilized for ratio-based metrics, such as tool execution success rates or Chain-of-Thought (CoT)~\cite{wei2023chainofthoughtpromptingelicitsreasoning} reasoning density. Let $v_{tgt}$ denote the target feature count and $v_{tot}$ the total feature count, with $M$ representing the maximum allocated score. Using the indicator function $\mathbb{I}$:
\begin{equation}
f_{ratio}(v_{tgt}, v_{tot}; M) = M \cdot \frac{v_{tgt}(\mathcal{T})}{v_{tot}(\mathcal{T})} \cdot \mathbb{I}\big(v_{tot}(\mathcal{T}) > 0\big)
\end{equation}

3. \textbf{Threshold Decay Penalty}
Designed to evaluate operational efficiency, such as interaction turns or token consumed. The agent receives maximum reward within an optimal range, beyond which the score decays linearly subject to a rigid lower bound. Let $c(\mathcal{T})$ denote the cost metric, $c_{min}$ the lower validity bound, $c_{opt}$ the optimal threshold, $p$ the decay penalty rate, and $m$ the minimum score limit:

\begin{equation}
\small
\begin{aligned}
    &f_{decay}(c; c_{min}, c_{opt}, p, M, m) = \\
    &\begin{cases}
    0, & c(\mathcal{T}) < c_{min} \\
    M, & c_{min} \le c(\mathcal{T}) \le c_{opt} \\
    \max \big( M - p \cdot (c(\mathcal{T}) - c_{opt}), m \big), & c(\mathcal{T}) > c_{opt}
    \end{cases}
\end{aligned}
\end{equation}

Given this generalized formulation, the proposed evaluation framework generalizes seamlessly across diverse agentic tasks. A Feature Extraction Agent extends this formulation by discovering task-specific features and adapting the scoring function $f_i$, together with reconfiguring the hyperparameter space $\Theta^{(\mathcal{S})}$, thereby achieving high scalability and domain-agnostic robustness.

\subsubsection{Macro-Level: Agentic Feature Extraction and Composition}

While the generalized scoring functions $f_i(\mathcal{T}; \theta_i)$ provide a robust mathematical foundation, manually configuring the hyperparameter space $\Theta^{(\mathcal{S})}$ for diverse scenarios relies heavily on human heuristics. To transcend this limitation, we introduce a closed-loop optimization pipeline combining an LLM-as-a-judge-based Feature Extraction Agent with Logistic Regression to discover and weight trajectory features.

We deploy a specialized Feature Extraction Agent, denoted as $\mathcal{E}$, which iteratively analyzes historical trajectory data to propose, evaluate, and refine feature formulations. The agent continuously contrasts fitting correlations to construct an $n$-dimensional optimal feature vector $\mathbf{x} = (x_1, x_2, \dots, x_n) \in \mathbb{R}^n$ from the raw trajectory $\mathcal{T}$.

The empirically discovered feature space $\mathcal{X}$ is categorized into four primary dimensions:

1. \textbf{Code Production ($\mathcal{X}_{code}$)} Quantifies tangible output, including the volume of modifications ($x_{lines\*changed}$) and the frequency of I/O interactions ($x*{file\_ops}$).

2. \textbf{Tool Usage ($\mathcal{X}_{tool}$)} Evaluates agentic capability. Beyond simple counts like total tool invocations ($x_{tool\*calls}$) and interaction rounds ($x*{agent\*turns}$), the agent constructs composite heuristics. For instance, the tool success rate ($x*{tool\*success}$) is parameterized conditionally, integrating temporal factors such as early-stage success rates and consecutive failure penalties. The breadth of capability is captured by tool diversity ($x*{tool\_diversity}$).

3. \textbf{Efficiency ($\mathcal{X}_{eff}$)} Measures the computational cost, primarily represented by the total token consumption ($x_{total\_tokens}$).

4. \textbf{Error Recovery ($\mathcal{X}_{recov}$)} Assesses robustness by counting active self-correction attempts ($x_{recovery\_attempts}$).

Through this agentic process, complex non-linear behaviors within the trajectory are mapped into discrete, evaluable features.

\subsubsection{Weight Fitting}

To determine the optimal contribution of each extracted feature to the final trajectory score, we model the evaluation as a binary classification problem. Let $y \in \{0, 1\}$ denote the ground-truth label of a trajectory, where $y=1$ represents a successful execution. We define the success criterion based on a thresholded reward signal:
\begin{equation}
y^{(j)} = \mathbb{I}\left(reward^{(j)} > 0.5\right)
\end{equation}
where $\mathbb{I}(\cdot)$ is the indicator function, and $reward^{(j)}$ is the original holistic reward for the $j$-th sample in the dataset.

Given the standardized feature vector $\tilde{\mathbf{x}}$ where each feature $x_i$ undergoes Z-score normalization and the weight vector $\mathbf{w} = (w_1, w_2, \dots, w_n)^T$, the decision function of the Logistic Regression model is defined as:
\begin{equation}
z = \mathbf{w}^T \tilde{\mathbf{x}} + b = \sum_{i=1}^{n} w_i \tilde{x}_i + b
\end{equation}

The probability of a trajectory being classified as successful is given by the sigmoid activation function. To find the optimal weight parameters $\mathbf{w}^*$, we optimize the model using the binary cross-entropy loss over $m$ trajectory samples.

\subsubsection{Integration with the Generalized Framework}

The Macro-Level of STITCH fitting pipeline consists of sequential stages: data loading, agentic feature extraction, binary label definition, feature standardization, and model training using bounded optimization (e.g., L2-regularized LR), followed by a validation stage.

In the verification stage, predictive performance is evaluated using standard metrics (e.g., accuracy and F1 score), and feature reliability is further validated via correlation analysis with respect to target outcomes. This dual validation mechanism ensures both the generalization capability of the model and the statistical significance of the extracted features.

Once the model converges, the extracted optimal coefficients $\mathbf{w}^*$ directly inform the hyperparameter space $\Theta^{(\mathcal{S})}$ of our generalized scoring framework. Specifically, the regression weight $w_i^*$ assigned to an extracted feature $x_i$ serves as the empirical basis for the linear weight $w$ or the slope penalty $p$ in the previously defined $f_{cap}$ and $f_{decay}$ functions. This data-driven mapping ensures that the evaluation framework dynamically adapts to the statistical reality of the agent's performance distribution.

\subsection{Micro-Level: Trajectory Semantic Segmentation and Analysis}

Evaluating extended Agent trajectories poses a significant challenge due to context windows limitations. To systematically assess trajectory quality without fragmenting the semantic coherence, we propose an auxiliary Trajectory Analysis Agent to address these constraints.

Unlike naive chunking methods that inevitably sever the logical chain between actions and observations, STITCH's Trajectory Analysis Agent reconstructs the evaluation context back together through a context-aware Map-Reduce paradigm coupled with a sliding memory mechanism.

Our approach decomposes the evaluation process into three integral phases: heuristic trajectory partitioning, context-aware local mapping, and global reduction.

\subsubsection{Trajectory Partitioning with Heuristic Safe-Split}

Formally, let an Agent's complete trajectory of length $N$ be defined as a sequence of steps $T = \{s_1, s_2, \dots, s_N\}$, where each step $s_t$ encapsulates the role, thought, action, or observation at time $t$. To prevent context overflow, $T$ must be partitioned into $K$ contiguous batches $\mathcal{B} = \{B_1, B_2, \dots, B_K\}$.

Naive fixed-length chunking inevitably disrupts atomic semantic boundaries (e.g., severing an Action from its corresponding environment Observation), leading to hallucinated or penalized evaluations. To mitigate this, we introduce a heuristic safety function $\sigma(s_t) \in \{0, 1\}$, where $\sigma(s_t) = 1$ indicates a safe split point (e.g., post-observation or upon receiving a new user instruction), and $\sigma(s_t) = 0$ strictly prohibits splitting (e.g., pending tool execution).

Subject to a minimum window size $L_{\min}$ and a maximum window size $L_{\max}$, the boundary index $t_k$ for the $k$-th batch is defined as:
\begin{equation}
t_k = \min \big( \mathcal{S}_k \cup \{ t_{k-1} + L_{\max} \} \big)
\end{equation}
where $\mathcal{S}_k = \{ t > t_{k-1} \mid t - t_{k-1} \ge L_{\min}, \sigma(s_t) = 1 \}$ represents the set of safe split points within the current window, $t_0 = 0$ and $t_K = N$. 

Consequently, the $k$-th observation batch is given by:

\begin{equation}
B_k = T[t_{k-1}+1 : t_k] = \{ s_{t_{k-1}+1}, \dots, s_{t_k} \}
\end{equation}

\subsubsection{Context-Aware Map Phase}

In the Map phase, we applied another LLM-as-a-judge-based method, denoted as the mapping function $f_{\text{map}}$, to perform local semantic segmentation and quality scoring on each batch $B_k$. 

To preserve chronological coherence and prevent semantic fragmentation across batch boundaries, we implement a Sliding Memory Mechanism. Let $m_k$ denote the last state summary generated at the end of batch $k$, with the initial state $m_0 = \emptyset$. For each batch, the evaluator receives both the current trajectory segment $B_k$ and the preceding memory state $m_{k-1}$. The mapping function outputs a set of scored semantic segments $S_k$ alongside the updated memory state $m_k$:

\begin{equation}
(S_k, m_k) = f_{\text{map}}(B_k, m_{k-1})
\end{equation}

The extracted segment set $S_k$ consists of $J_k$ granular sub-tasks, represented as:
\begin{equation}
S_k = \big\{ (c_1, q_1), (c_2, q_2), \dots, (c_{J_k}, q_{J_k}) \big\}
\end{equation}

where $c_j$ encapsulates the semantic summarization of the $j$-th segment (including start/end indices and structural intent), and $q_j \in [1, 10]$ is the corresponding quality score evaluating the Agent's local reasoning and execution efficiency.

\subsubsection{Global Reduce Phase}

The Map phase yields a sequence of highly condensed, scored segments that abstract away redundant token-level details. In the Reduce phase, we aggregate these local segments into a unified, abstracted trajectory representation:

\begin{equation}
T_{\text{abstract}} = \bigcup_{k=1}^K S_k
\end{equation}

Finally, a global reduction function $f_{\text{reduce}}$ is applied to $T_{\text{abstract}}$ to evaluate the Agent's macroscopic behavior. This includes assessing strategic coherence, task completion rates, and the absence of systemic loops. The overall evaluation $E_{\text{global}}$ is formulated as:

\begin{equation}
E_{\text{global}} = f_{\text{reduce}}\left( T_{\text{abstract}} \right)
\end{equation}

\subsubsection{Summary of the Micro-Level Framework}

By integrating the sliding memory state into the Map-Reduce paradigm, the Micro-Level part of STITCH ensures that local contextual awareness is maintained without sacrificing computational scalability. The entire algorithmic workflow can be elegantly summarized by:

\begin{equation}
E_{\text{global}} = f_{\text{reduce}} \left( \bigcup_{k=1}^K \text{proj}_{S} \Big( f_{\text{map}}(B_k, m_{k-1}) \Big) \right)
\end{equation}

where $\text{proj}_{S}$ denotes the projection operator that extracts the segment set $S_k$ from the tuple $(S_k, m_k)$ while propagating the memory state $m_k$ forward.  Given complete trajectory data, this approach further filters out important segments of the trajectory that represent complete behavior.

\subsection{Two-Stage Trajectory Curation Pipeline} 

To construct high-quality datasets for downstream finetuning task while maintaining computational efficiency, we integrate the data-driven LR model with the previously proposed STITCH framework. This integration forms a robust two-stage selection pipeline: Macro-Level Pre-screening and Micro-Level Semantic Extraction.

Let the initial raw dataset of Agent trajectories be denoted as $\mathcal{D}_{\text{raw}} = \{\mathcal{T}_1, \mathcal{T}_2, \dots, \mathcal{T}_M\}$.

\subsubsection{Macro-Level Pre-screening via Statistical Features}
In the first stage, we leverage the computationally lightweight LR model, parameterized by the optimal weights $\mathbf{w}^*$, to act as a global filter. For each raw trajectory $\mathcal{T} \in \mathcal{D}_{\text{raw}}$, the agentic feature extraction pipeline derives the standardized feature vector $\tilde{\mathbf{x}}$. We compute the success probability $P(y=1|\tilde{\mathbf{x}}) = \sigma({\mathbf{w}^*}^T \tilde{\mathbf{x}} + b)$. 

Trajectories that fail to meet a globally defined heuristic confidence threshold $\tau_{\text{global}}$ are discarded. This efficiently prunes systemic failures, meaningless loops, and low-reward generations. The pre-screened candidate pool $\mathcal{D}_{\text{cand}}$ is formally defined as:
\begin{equation}
\mathcal{D}_{\text{cand}} = \Big\{ \mathcal{T} \in \mathcal{D}_{\text{raw}} \;\Big|\; \sigma({\mathbf{w}^*}^T \tilde{\mathbf{x}} + b) \ge \tau_{\text{global}} \Big\}
\end{equation}
This coarse-grained filtering significantly reduces the token consumption and computational overhead required for the subsequent LLM-as-a-judge evaluation.

\subsubsection{Micro-Level Semantic Verification and Extraction}
While the LR model effectively identifies statistically promising trajectories, it lacks the semantic depth to pinpoint where the Agent performed well or to detect subtle logical hallucinations. Therefore, in the second stage, we apply the Trajectory Analysis Agent to the candidate pool $\mathcal{D}_{\text{cand}}$.

For each candidate trajectory $\mathcal{T} \in \mathcal{D}_{\text{cand}}$,  Trajectory Analysis Agent maps the sequence into a set of scored semantic segments $T_{\text{abstract}} = \bigcup_{k=1}^K S_k$, where each segment tuple $(c_j, q_j)$ contains the sub-task context $c_j$ and its localized quality score $q_j$. Furthermore, Trajectory Analysis Agent conducts a global reduction to yield the definitive semantic evaluation $E_{\text{global}}$.

This fine-grained stage serves multi purposes depending on the downstream alignment methodology. In our case, besides using the entire trajectory, we also isolate only the rigorously executed sub-tasks. By setting a localized segment threshold $\tau_{\text{seg}}$, we curate a pristine SFT dataset $\mathcal{D}_{\text{SFT}}$ consisting of optimal action-observation pairs:
\begin{equation}
\mathcal{D}_{\text{SFT}} = \left\{ c_j \mid (c_j, q_j) \in T_{\text{abstract}} \right\}
\end{equation}
where each sample must satisfy $q_j \ge \tau_{\text{seg}}$ for all $\mathcal{T} \in \mathcal{D}_{\text{cand}}$.

Even with trajectory data that scores low from a global perspective, this method can still extract valuable behavioral fragments, allowing for greater utilization of existing data.

By combining the macroscopic statistical bounds of Logistic Regression with the microscopic semantic rigor of STITCH, this two-stage pipeline guarantees that the resulting datasets are both statistically robust and semantically flawless, fundamentally accelerating the downstream alignment of LLM-based Agents.

\section{Experimental Setup}

\subsection{Experimental Settings}
We use the same settings on different models ranging from 30B to 230B. We use MindSpeed-LLM \cite{mindspeedllm} as our training framework. Models are trained for 4 epochs with a global batch size of 64 on 64 to 256 Ascend 910C NPUs. The maximum sequence length is set to 81920. We use AdamW optimizer \cite{adamW} and cosine learning rate schedule with a warmup ratio of 0.05. The maximum learning rate is set to 5e-6 and decays to 5e-7 as a minimum. During evaluating, we set temperature to 0.7 for all experiments and evaluate three times to calculate an average.

We collect real GitHub issues from popular repositories and make sure that issues related to testsets are decontaminated. In the end, We use 752, 391 and 924 high quality data for Python, Java, and ArkTS, respectively. All data is constructed by our SandForge framework and filtered by our STITCH. Only tokens generated by models and judged as high value are included in loss calculation.

\subsection{Evaluation}
For different programming languages and scenarios, we use various benchmarks to evaluate model's agent capability. We use SWE-bench-Verified \cite{SWE-bench}, the most widely used benchmark in the field, to evaluate the model's capability in Python SWE scenarios. For Java scenarios, we employ Mulit-SWE-bench(Java) \cite{multi-swe-bench}, a subset of Multi-SWE-bench which contains 128 real GitHub issues collected from 9 different repositories. For ArkTS, we construct a benchmark which consists of 168 requirements from real HarmonyOS developers and aims at evaluating model's capability of creating a HarmonyOS application from scratch.

\subsubsection{SWE-bench-Verified(Python) Setup}

We use mini-SWE-agent \cite{minisweagent} to run the trajactories of all 500 cases. Mini-SWE-agent only allow models to generate bash lines to complete tasks and only one bash line is allowed in one response. We set the limitation of chat turns to 200 for all models and run three times at the same temperature(0.7) to calculate the average.

\subsubsection{Multi-SWE-Bench(Java) Setup}

We evaluate with \texttt{MSWE-agent} and our internal \texttt{CodeArts-Agent}. The \texttt{MSWE-agent} is used to measure the performance of the trained model in the reference agent stack of the benchmark. \texttt{CodeArts-Agent} is used to assess the trained model when paired with a production-grade coding agent, reflecting deployment-relevant capability beyond the reference configuration. We set the maximum number of chat turns to 200 for \texttt{MSWE-agent} and 500 for \texttt{CodeArts-Agent}, and use the same temperature of 0.7 for all models. In addition to automated runs through the \textbf{SandForge} framework, we deploy an evaluation environment that is fully equivalent to the official \textbf{Multi-SWE-Bench} setup, in order to isolate any effects due to implementation or integration differences. The benchmark results reported in this article are obtained in this equivalent official evaluation environment.

\subsubsection{HarmonyOS(ArkTS) Evaluation Setup}
We evaluate model performance on the 168-problem test set using two complementary metrics that progressively measure the correctness of the model's outputs:

\paragraph{Compilation Pass Rate.} The percentage of test problems for which the model generates a project that successfully passes the ArkTS compiler without errors. This metric reflects the model's ability to produce syntactically valid and type-correct code under ArkTS's strict static type system, serving as a necessary condition for functional correctness.

\paragraph{Preview Pass Rate.} The percentage of test problems for which the model generates a project that not only compiles successfully but also implements the required functionality. For each compilable project, we deploy it to a HarmonyOS device via the HDC toolchain, capture screenshots of all UI pages, and use a multimodal evaluation model (Qwen/Qwen3.5-122B-A10B) to judge whether the rendered interface matches the original requirement description. This metric is computed over the entire test set (not just compilable projects), providing a holistic measure of end-to-end model capability—from code generation to functional correctness.

\subsubsection{Evaluation Integrity and Leakage Prevention}

Because SWE benchmarks are often derived from real open-source repositories, they may contain enough contextual information to leak solution signals. We therefore apply explicit evaluation controls to reduce answer leakage during benchmark runs.

First, agents are not allowed to use \textbf{Web Fetch} or equivalent Internet-retrieval tools. We acknowledge that some benchmark instances include issue URLs, image links, or similar references inside the issue description, and disabling web access may therefore prevent the agent from fully recovering all issue context. Nevertheless, we adopt a strict no-web-fetch policy because the leakage risk from unrestricted external retrieval outweighs the potential benefit of recovering additional context.

Second, we disallow history-oriented Git inspection paths such as \textbf{git log} and \textbf{git show}, which may directly expose future commits, patches, or repository states closely related to the target repair. For agents that support direct configuration-based restrictions, we disable these capabilities in the agent configuration. For agents that cannot fully disable them through configuration alone, we perform post-run trajectory matching and mark any otherwise successful run as failed if its trace contains either of these commands.

\section{Result}


\begin{table*}[!ht]
  \centering
  \resizebox{\linewidth}{!}{%
  \begin{tabular}{l l c r c}
    \toprule
    Model & Scaffold & Data Source & Resolve Rate (\%) & \makecell{Relative\\Improvement (\%)} \\
    \midrule

    \multicolumn{5}{c}{\textbf{Proprietary Models}} \\
    \midrule
    Claude 4.5 Opus medium (2025-11-01) & live-SWE-agent & Official Leaderboard & 79.20 & - \\
    Gemini 3 Pro Preview (2025-11-18) & live-SWE-agent & Official Tested & 77.40 & - \\
    Claude 4.5 Opus (high reasoning) & mini-SWE-agent & Official Leaderboard & 76.80 & - \\
    Gemini 3 Flash (high reasoning) & mini-SWE-agent & Official Leaderboard & 75.80 & - \\
    GPT-5-2 Codex & mini-SWE-agent & Official Leaderboard & 72.80 & - \\

    \midrule
    \multicolumn{5}{c}{\textbf{Open-Source Models}} \\
    \midrule

    \rowcolor{gray!20}
    \multicolumn{5}{l}{\textit{Parameters $\approx$ 30B}} \\
    \midrule
    Qwen2.5-Coder 32B Instruct & mini-SWE-agent & Official Leaderboard & 9.00 & - \\
    Qwen3-Coder-30B-A3B-Instruct & mini-SWE-agent & Self Test & 26.60 & - \\
    \rowcolor{cyan!10}
    Qwen3-Coder-30B-A3B-Instruct-RFT (ours) & mini-SWE-agent & Self Test & 34.40 & +29.32\% \\
    \rowcolor{cyan!10}
    Qwen3-Coder-30B-A3B-Instruct-STITCH (ours) & mini-SWE-agent & Self Test & 43.40 & +63.16\% \\

    \midrule
    \rowcolor{gray!20}
    \multicolumn{5}{l}{\textit{Parameters $\approx$ 100B}} \\
    \midrule
    GLM-4.5-Air (106B-A12B) & mini-SWE-agent & Self Test & 42.20 & - \\
    \rowcolor{cyan!10}
    GLM-4.5-Air-RFT (106B-A12B, ours) & mini-SWE-agent & Self Test & 43.00 & +1.90\% \\
    \rowcolor{cyan!10}
    GLM-4.5-Air-STITCH (106B-A12B, ours) & mini-SWE-agent & Self Test & 48.40 & +14.69\% \\

    \midrule
    \rowcolor{gray!20}
    \multicolumn{5}{l}{\textit{Parameters $>$ 200B}} \\
    \midrule
    MiniMax M2.5 (high reasoning, 230B-A10B) & mini-SWE-agent & Official Leaderboard & 75.80 & - \\
    MiniMax M2.5 (230B-A10B) & mini-SWE-agent & Self Test & 73.80 & - \\
    \rowcolor{cyan!10}
    MiniMax M2.5-RFT (230B-A10B, ours) & mini-SWE-agent & Self Test & 72.80 & -1.36\% \\
    \rowcolor{cyan!10}
    MiniMax M2.5-STITCH (230B-A10B, ours) & mini-SWE-agent & Self Test & 75.80 & +2.71\% \\
    GLM-4.7 (355B-A32B) & mini-SWE-agent & Self Test & 66.80 & - \\
    \rowcolor{cyan!10}
    GLM-4.7-RFT (355B-A32B, ours) & mini-SWE-agent & Self Test & 64.80 & -2.99\% \\
    \rowcolor{cyan!10}
    GLM-4.7-STITCH (355B-A32B, ours) & mini-SWE-agent & Self Test & 68.40 & +2.39\% \\

    \bottomrule
  \end{tabular}
  }
  \caption{Model Performance on SWE-bench Verified}
  \label{tab:model_performance}
\end{table*}

\begin{table*}[!ht]
  \centering
  \resizebox{\linewidth}{!}{%
  \begin{tabular}{l l c r c}
    \toprule
    Model & Scaffold & Data Source & Resolve Rate (\%) & \makecell{Relative\\Improvement (\%)} \\
    \midrule

    \multicolumn{5}{c}{\textbf{Proprietary Models}} \\
    \midrule
    GPT5.2 & InfCode & Official Leaderboard & 39.06 & - \\
    Gemini-2.5-Pro & MSWE-agent & Official Tested & 28.91 & - \\
    Claude-3.7-Sonnet & MSWE-agent & Official Tested & 23.44 & - \\

    \midrule
    \multicolumn{5}{c}{\textbf{Open-Source Models}} \\
    \midrule

    \rowcolor{gray!20}
    \multicolumn{5}{l}{\textit{Parameters $>$ 200B}} \\
    \midrule
    DeepSeek-R1 (671B-A37B) & MagentLess & Official Tested & 22.65 & - \\
    MiniMax-M2.5 (230B-A10B) & MSWE-agent & Self Test & 22.65 & - \\
    
    \rowcolor{cyan!10}
    MiniMax-M2.5-RFT (230B-A10B, ours) & MSWE-agent & Self Test & 22.65 & +0.00\% \\
    
    \rowcolor{cyan!10}
    MiniMax-M2.5-STITCH (230B-A10B, ours) & MSWE-agent & Self Test & 27.34 & +20.71\% \\
    
    MiniMax-M2.5 (230B-A10B) & CodeArts Agent & Self Test & 37.50 & - \\
    
    \rowcolor{cyan!10}
    MiniMax-M2.5-RFT (230B-A10B, ours) & CodeArts Agent & Self Test & 38.28 & +2.08\% \\
    
    \rowcolor{cyan!10}
    MiniMax-M2.5-STITCH (230B-A10B, ours) & CodeArts Agent & Self Test & 43.75 & +16.67\% \\
    \bottomrule
  \end{tabular}
  }
  \caption{Model Performance on Multi-SWE-bench(Java)}
  \label{tab:model_performance_java}
\end{table*}

\begin{table*}[!ht]
  \centering
  \resizebox{\linewidth}{!}{%
  \begin{tabular}{l l c cc cc}
    \toprule
    \multirow{2}{*}{Model} & \multirow{2}{*}{Scaffold} & \multirow{2}{*}{Data Source} 
    & \multicolumn{2}{c}{Score (\%)} & \multicolumn{2}{c}{Relative Improvement (\%)} \\
    \cmidrule(lr){4-5} \cmidrule(lr){6-7}
    & & & Compile & Render & Compile Imp. & Render Imp. \\
    \midrule

    GLM-4.7 (355B-A32B) & CodeArts Agent & Self Test & 42.77 & 31.54 & - & - \\

    \rowcolor{cyan!10}
    GLM-4.7-STITCH (355B-A32B, ours) & CodeArts Agent & Self Test & 61.31 & 44.05 & +43.34\% & +39.66\% \\

    \bottomrule
  \end{tabular}
  }
  \caption{Model Performance on HarmonyOS (ArkTS)}
  \label{tab:model_performance_arkts}
\end{table*}

\subsection{Python Results}

We conducted experiments on models with sizes from 30B to 355B to validate our approach on Python. As shown in Table~\ref{tab:model_performance}, compared with vanilla Reject Sampling Fine Tuning (RFT), training data filtered by STITCH shows better effectiveness across all model sizes. The gap is most pronounced on small-scale models (30B). After training with the STITCH method, the probability that the model generates runnable bash lines is significantly improved, and the probability of successfully completing and submitting tasks is also increased. For medium-scale models ($\sim$100B), the original RFT method yields only limited improvements, yet STITCH filtering still delivers considerable gains. For large-scale SOTA models ($>$200B), although STITCH achieves better average performance, we do not observe highly significant improvements. We think that SOTA models might have approached saturation on this benchmark.
\subsection{Java Results}

Table~\ref{tab:model_performance_java} shows the results on Multi-SWE-Bench (Java), the STITCH method yields consistent improvements on both \texttt{MSWE-agent} and \texttt{CodeArts-Agent}. At the time of writing, the \texttt{CodeArts-Agent} result achieves state-of-the-art performance on the MSWE-bench(Java) leaderboard, surpassing approaches built on leading closed-source models such as GPT-5.2 and Claude-4.5-Sonnet.

From these results, we observe the following phenomena. First, coding agents based on the function-calling format (e.g., \texttt{CodeArts-Agent}) appear to have an advantage over ReAct-style agents with customized action formats (e.g., \texttt{MSWE-agent}) when paired with newer large-scale models. We hypothesize that this is because such models have been extensively trained for compliance with function-calling formats prior to release, and function-calling agents therefore align better with the models' built-in capabilities. Second, STITCH consistently outperforms RFT on both \texttt{CodeArts-Agent} and \texttt{MSWE-agent}. This finding supports our hypothesis that improving trajectory quality can further enhance training efficiency.

\subsection{ArkTS Results}

To evaluate the generalizability of our STITCH approach beyond high-resource languages such as Java, we conduct experiments on ArkTS—a statically-typed language developed by Huawei for HarmonyOS application development. ArkTS is derived from TypeScript with additional static type constraints, making it a representative low-resource language with limited publicly available training data. Demonstrating improvements on ArkTS validates that our method is not confined to well-resourced programming languages but generalizes effectively to emerging, underrepresented language ecosystems.

\subsubsection{Experimental Results}

Table~\ref{tab:model_performance_arkts} presents the evaluation results on the ArkTS benchmark. Several key observations can be drawn:

\paragraph{\mbox{Significant improvements with minimal data.}}

With only approximately 1K STITCH training trajectories, GLM-4.7 achieves an absolute improvement of 18.54 percentage points in Compilation Pass Rate (from 42.77\% to 61.31\%) and 12.51 percentage points in Preview Pass Rate (from 31.54\% to 44.05\%). This demonstrates that a small volume of high-quality, boundary-aware training data can drive substantial performance gains, consistent with the ``less is more'' philosophy of our approach.

\paragraph{\mbox{Generalization to low-resource languages.}}

ArkTS presents a particularly challenging setting due to the scarcity of publicly available training data and its divergence from TypeScript in static type constraints. The strong improvements observed here, combined with our results on Java (a high-resource language), confirm that the STITCH methodology generalizes effectively across languages with vastly different resource availability. The core mechanism—identifying and learning from model-boundary samples—is language-agnostic and remains effective even when the target language is underrepresented in pretraining corpora.

\paragraph{\mbox{Addressing systematic syntax biases.}}\leavevmode

\noindent A detailed analysis of compilation errors reveals that the majority of failures in the base model stem from violations of ArkTS-specific migration rules (error codes \texttt{10605XXX}), particularly the prohibited use of \texttt{any} types. After training, \texttt{any}-related errors decrease significantly, indicating that the model successfully learns to distinguish ArkTS's strict static type requirements from TypeScript's more permissive type system.

\subsubsection{Case Studies}

We present two representative cases illustrating the qualitative improvements achieved by training, from the perspectives of code correctness and visual quality, respectively.

\paragraph{Case 1: Code comparison.}

\begin{figure}[!ht]
\centering
\begin{minipage}[t]{0.48\linewidth}
  \centering
  \includegraphics[width=\linewidth]{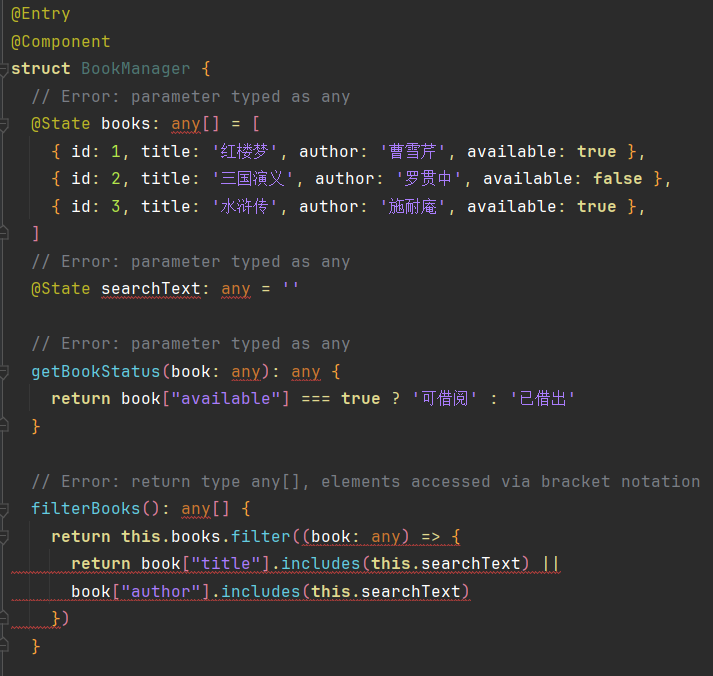}
\end{minipage}
\hfill
\begin{minipage}[t]{0.48\linewidth}
  \centering
  \includegraphics[width=\linewidth]{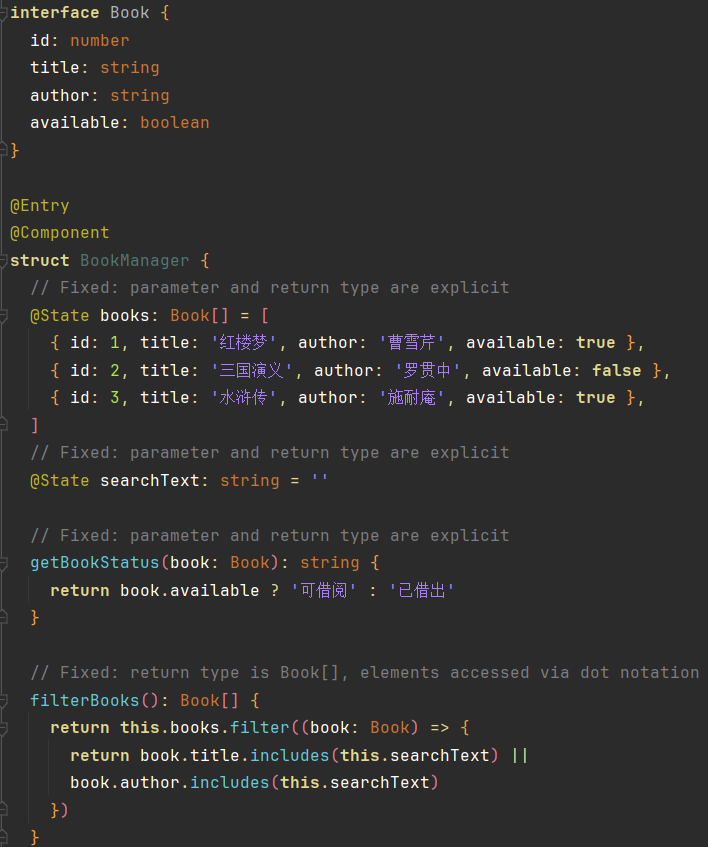}
\end{minipage}
\caption{Code comparison for the same requirement. \textbf{Left}: the base model generates TypeScript-style code with \texttt{any} types and literal object types, causing compilation failures. \textbf{Right}: the trained model produces ArkTS-compliant code with explicit type declarations that compiles successfully.}
\label{fig:case_compilation}
\end{figure}

Figure~\ref{fig:case_compilation} contrasts representative code snippets generated by the base model and the trained model for the same requirement. The base model produces code that relies heavily on \texttt{any} type annotations, literal object types, and dynamic property access via bracket notation (e.g., \texttt{obj["key"]})—patterns that are valid in TypeScript but violate ArkTS's strict static type constraints, resulting in compilation failures. After training, the model generates ArkTS-compliant code with explicit interface definitions, proper static type declarations, and dot-notation property access, which compiles successfully without errors.


\paragraph{Case 2: Rendering comparison.}

\begin{figure}[!ht]
\centering
\begin{minipage}[t]{0.35\linewidth}
  \centering
  \includegraphics[width=\linewidth]{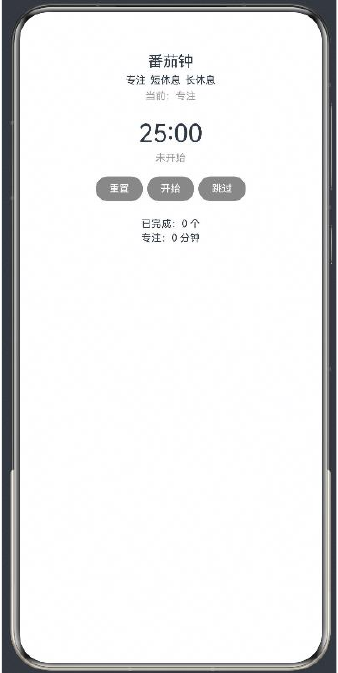}
\end{minipage}
\hfill
\begin{minipage}[t]{0.35\linewidth}
  \centering
  \includegraphics[width=\linewidth]{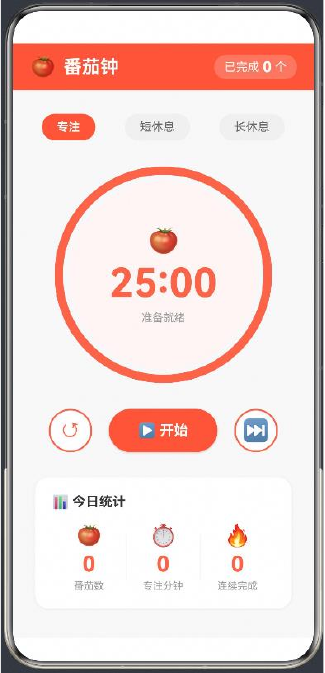}
\end{minipage}
\caption{Rendering comparison for the same requirement. \textbf{Left}: the base model generates a visually rough interface with improperly sized components and inconsistent styling. \textbf{Right}: the trained model produces a polished, well-laid-out interface matching the requirements.}
\label{fig:case_visual}
\end{figure}

Figure~\ref{fig:case_visual} compares the rendered UI screenshots of projects generated by the two models for the same requirement, where both projects pass compilation. The base model produces a visually rough interface: components are improperly sized, colors lack contrast and consistency, and the overall layout appears disorganized. In contrast, the trained model generates a polished interface with well-proportioned components, harmonious color schemes, and a coherent layout that faithfully reflects the requirement specification. This case demonstrates that training data filtered by STITCH improves not only code-level correctness but also the model's ability to produce aesthetically refined and functionally complete applications.



\section{Conclusion}

In this paper, we propose a "Less-Is-More"-style training framework, demonstrating that stronger agentic capabilities can be achieved with less but higher-quality training data. Central to our framework is STITCH, a coarse-to-fine trajectory curation mechanism that combines statistical pre-screening with LLM-based semantic analysis via a Map-Reduce paradigm with sliding memory, preserving cross-segment coherence and enabling high-value segment extraction even from globally suboptimal trajectories. 
Experiments across multiple agent frameworks, model scales (30B to 355B), and multilingual settings (Python, Java, and ArkTS) consistently confirm that the "Less-Is-More" paradigm extends naturally from mathematical reasoning to the broader domain of coding and agentic tasks.

\section*{Contributors \footnote{$^*$ Equal contribution, $^\dagger$ Corresponding author}}
\label{sec:contributors}
\begin{minipage}[t]{0.48\textwidth}
Yang Ye$^*$ \hfill \texttt{yeyang14@huawei.com} \\
Jingyuan Tan$^*$ \hfill \texttt{tanjingyuan2@huawei.com}\\
Tianyue Jiang$^*$ \hfill \texttt{jiangty9@mail2.sysu.edu.cn}\\
Ruizhe Ye$^*$ \hfill \texttt{yeruizhe@huawei.com}\\
Qiankun He \hfill \texttt{heqiankun1@huawei.com}\\
Jiarui Yang \hfill \texttt{yangjiarui@huawei.com}\\
Jian Dong \hfill \texttt{dongjian26@huawei.com}\\
Sicong Liang \hfill \texttt{liangsicong2@huawei.com}\\
Chongjian Yue \hfill \texttt{yuechongjian@huawei.com}\\
Peibai Xu \hfill \texttt{xupeibai1@huawei.com}\\
Lufan Lu \hfill \texttt{lulufan@huawei.com}\\
Shiguan Pang \hfill \texttt{pangshiguan1@huawei.com}\\
Taotao Qian \hfill \texttt{qiantaotao@huawei.com}\\
Junbao Hu \hfill \texttt{hujunbao1@huawei.com}\\
Yuechan Hao $^\dagger$ \hfill \texttt{haoyuechan@huawei.com}\\
Ensheng Shi \hfill \texttt{shiensheng@huawei.com}\\
Qi Zhang \hfill \texttt{Kinopico.Zhang@huawei.com}\\
Yi Hao \hfill \texttt{haoyi6@huawei.com}\\
Na Fan \hfill \texttt{fanna.fan@huawei.com}\\
Xin Tan \hfill \texttt{tanxin50@huawei.com}\\
Shuai Yao \hfill \texttt{yaoshuai1@huawei.com}\\
Zhiwei Shen \hfill \texttt{shenzhiwei5@huawei.com}\\
Zongchen Li \hfill \texttt{lizongchen@huawei.com}\\
Yanlin Wang \hfill \texttt{wangylin36@mail.sysu.edu.com}\\
Chong Chen \hfill \texttt{chenchong55@huawei.com}\\
Yuchi Ma \hfill \texttt{mayuchi1@huawei.com}\\
\end{minipage}\hfill%


\bibliography{custom}
\clearpage
\onecolumn
\section{Appendix}
\label{sec:appendix}

\begin{figure}[!ht]
  \centering
  \includegraphics[width=1\linewidth]{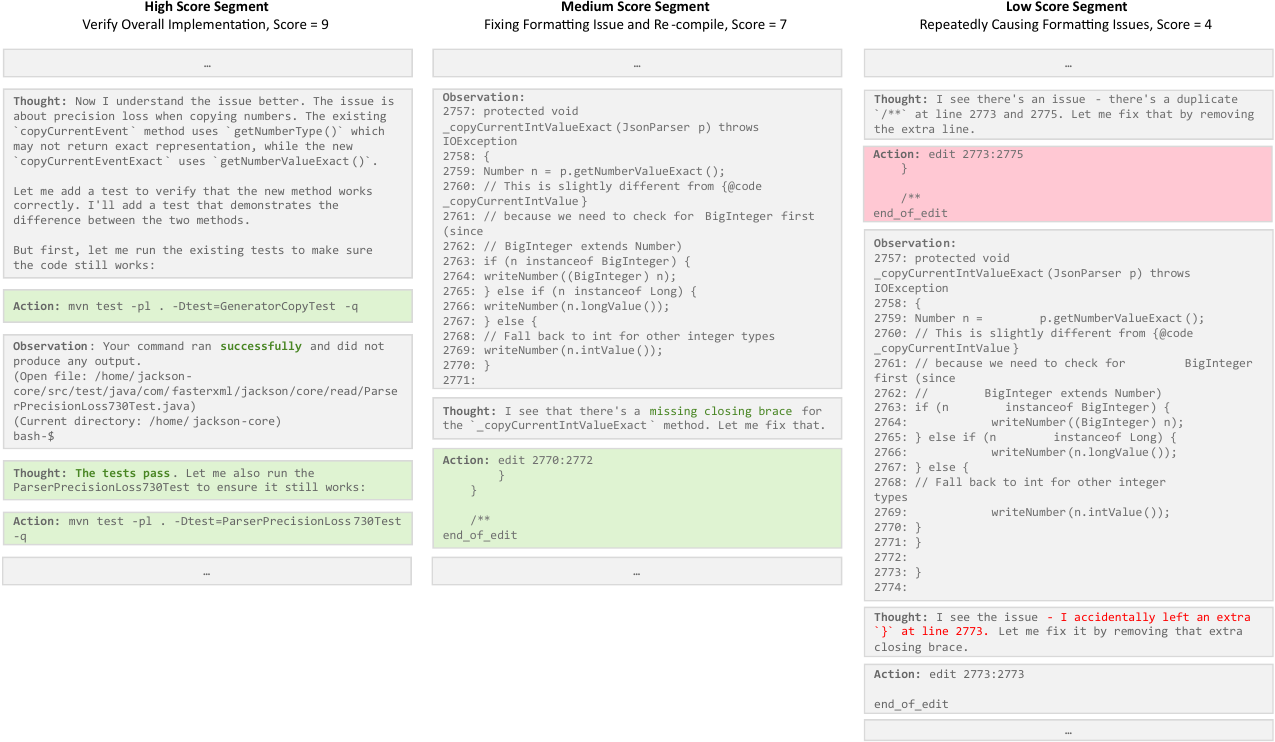}
  \caption{Scored Segment Sample from Curated Trajectories \textbf{Left}: High-scoring segments where writing and test execution steps are correctly and seamlessly integrated. \textbf{Middle}: Agent identified problems in the preceding results accurately resolved them after reflection being recognized as medium score segment. \textbf{Right}: Agent failed to pinpoint the root cause of the problem in the first attempt, repeatedly introducing new and unnecessary issues, which is a typical low score segment.}
  \label{fig:STITCH_case_combined}
\end{figure}

\end{document}